# Topological Nature of Radiation Asymmetry in Bilayer Metagratings

Ze-Peng Zhuang, Hao-Long Zeng, Xiao-Dong Chen, Xin-Tao He and Jian-Wen Dong*

School of Physics & State Key Laboratory of Optoelectronic Materials and Technologies, Sun Yat-sen University, Guangzhou 510275, China

*Corresponding author: dongjwen@mail.sysu.edu.cn

**Abstract:**

Manipulating radiation asymmetry of photonic structures is of particular interest in many photonic applications such as directional optical antenna, high efficiency on-chip lasers, and coherent light control. Here, we proposed a term of pseudo-polarization to reveal topological nature of radiation asymmetry in bilayer metagratings. Robust pseudo-polarization vortex with an integer topological charge exists in P-symmetry metagrating, allowing for tunable directionality ranging from -1 to 1 in synthetic parameter space. When P-symmetry-breaking, such vortex becomes pairs of C points due to the conservation law of charge, leading to the phase difference of radiation asymmetry from $\pi/2$ to $3\pi/2$. Furthermore, topologically enabled coherent perfect absorption is robust with customized phase difference at will between two counter-propagating external light sources. This work can not only enrich the understanding of two particular topological photonic behavriors, i.e., bound state in the continuum and unidirectional guided resonance, but also provide a topological view on radiation asymmetry, opening an unexplored avenue for asymmetric light manipulation in on-chip laser, light-light switch and quantum emitters.

**Introduction**

Radiation asymmetry (RA) in upward and downward directions is highly desirable for high-efficiency energy utilization and asymmetric light manipulation in photonic devices. Large amount of works have been dedicated to enhancing the radiation directionality, such as Yagi-Uda antennas [1-3], Mie scatterers under the Kerker condition [4-6], on-chip lasers [7-9] and grating couplers [10-12] in photonic integrated circuits. In contrast, another aspect of RA, that is radiation phase difference, is always ignored. This hinders phase as an additional degree of freedom in coherent light control including coherent absorption [13-15], coherent polarization control [16,17] and light-light switching [18-20]. In general, arbitrary control of RA including directionality and phase difference is significant for light field manipulation but remains elusive.

On the other hand, topological perspective has been employed to manipulate radiation field of photonic structures recently [21-23]. By modulating polarization vortexes in momentum space, bound states in the continuum (BICs) and unidirectional guided resonances (UGRs) with distinct radiation properties have been confirmed. Although lying in the continuum spectrum with potential radiation channels, BIC manifests itself as a radiation singularity without leaky energy. It is situated at the polarization vortex center (V point) carrying an integer topological charge [24-26]. The topological property of BICs not only facilitates the design of robust and high-Q resonators [27-30], but also introduces a new paradigm for polarization manipulation [31-33]. Intriguing applications based on BICs, such as lasing [34-37], chirality detection [38-40] and resonant wavefront shaping [41-43], have been demonstrated. Unlike BICs, UGRs radiate only in a single channel. It can be interpreted as the generation of circular polarized states (C points) from the broken V points [44] or voids [45,46], followed by the merging of C points in a single channel while leaving them apart in the other channel. Although BICs and UGRs have been demonstrated as two types of special RA, both of them have ignored the radiation phase difference. What's more, the evolution of polarization vortex in momentum space fails to describe situations involving arbitrary directoinality and phase difference that fall in between BICs and UGRs.

In this letter, taking both directionality and phase difference into account, we propose a RA pseudo-polarization. By mapping the RA pseudo-polarization onto the Poincaré sphere, a unified physical picture is developed to describe arbitrary RA, including radiation behaviors of BICs and UGRs. We find that a pseudo-polarization vortex resulting from BIC robustly exists in the synthetic parameter space of bilayer metagratings with an integer topological charge. The pseudo-polarization vortex obeys the conservation law of topological charge and can be broken into a pair of C points, revealing the topological nature of RA. Through manipulation of the vortex and C points, we further validate the realization of flexible RA including directionality ranging from -1 to 1 and phase difference from $\pi/2$ to $3\pi/2$. Finally, we demonstrate that the ability of controlling radiation phase difference induced by C points enables the design of coherent perfect absorption under customized phase difference of input light.

**Results**

Generally, the up-down mirror-symmetry-breaking structures, for example, photonic crystal slabs, can support leaky modes that exhibit asymmetric radiation in upward and downward directions. As shown in Fig. 1(a), the complex amplitude of radiated plane wave is denoted as $c_{up}$ and $c_{down}$ for upward and downward direction, respectively. We introduce a pseudo-polarization $\vec{c}$, which we name radiation asymmetry (RA) pseudo-polarization, as a measure of radiation difference between these two directions. The RA pseudo-polarization is defined as

$$\vec{c} = \frac{1}{N}\left[\left(c_{up} + c_{down}\right)\vec{e}_1 + \left(c_{up} - c_{down}\right)\vec{e}_2\right] \tag{1}$$

where $\vec{e}_1$ and $\vec{e}_2$ are a pair of orthogonal basis. $N = \sqrt{\left|c_{up} + c_{down}\right|^2 + \left|c_{up} - c_{down}\right|^2}$ is the normalization coefficient. Note that the first term ($c_{up} + c_{down}$) represents the symmetric component of radiation fields and when both upward and downward radiation fields have identical amplitudes and phases, this component reaches the maximum. Similarly, ($c_{up} - c_{down}$) represents the anti-symmetric component. Generally, the amplitude and phase of $c_{up}$ are different from $c_{down}$. As a consequence, the RA pseudo-polarization is generally a pseudo-polarization ellipse with orientation angle $\psi$ and ellipticity angle $\chi$.

By mapping the RA pseudo-polarization ellipse onto the Poincaré sphere, we can effectively characterize the RA in a clear physical picture. As shown in Fig. 1(b), each point on the Poincaré sphere corresponds to a specific pseudo-polarization ellipse as well as relating to a kind of RA. More quantitatively, we define directionality as $\eta = \frac{\left|c_{up}\right|^2 - \left|c_{down}\right|^2}{\left|c_{up}\right|^2 + \left|c_{down}\right|^2}$ and phase difference as $\delta = \arg\left(\frac{c_{up}}{c_{down}}\right)$, which respectively characterize the radiation amplitude and phase difference. After some derivation (Supplemental Materials [47], Sec. S2.), these two parameters are actually related to Stokes parameters of Poincaré sphere as,

$$S_1 = \sqrt{1-\eta^2}\cos\delta$$
$$S_2 = \eta \quad\quad (2)$$
$$S_3 = \sqrt{1-\eta^2}\sin\delta$$

Eq. (2) clearly shows that directionality $\eta$ is the projection length of Stokes vector on $S_2$ axis (double-headed arrow in Fig. 1(c)) and phase difference $\delta$ refers to the angle between $S_1$ axis and ($S_1$, $S_3$) in-plane component of Stokes vector (arc in Fig. 1(c)).

General RA with various directionality and phase difference can be described by Eq. (2), and are supposed to cover the whole Poincaré sphere. Specially, two poles of $S_2$ axis represents 45° or 135° linear pseudo-polarization and they correspond to unidirectional radiation in the downward or upward direction, which exactly describes the radiation behavior of unidirectional guided resonance (UGR, red dots in Fig. 1(c)). Moreover, the center of Poincaré sphere implies a singular case, where the phase difference is undefined but the amplitudes of upward and downward radiation fields are equal. This singular case actually corresponds to bound state in the continuum (BIC) since both upward and downward radiation fields are zero.

The RA pseudo-polarization then helps us to investigate the RA of photonic structures and unveil its topological nature in the parameter space. As a specific example, we consider a bilayer metagratings as shown in Fig. 2(a). The bilayer metagratings is composed of one-dimensional silicon (n = 3.4767) gratings covered in SiO$_2$ (n = 1.444) background with G = 0.045a, $w_g$ = $w_e$ =0.704a and $t_g$ = 0.355a, where a is the lattice length. By introducing a nonzero lateral offset Δg, the up-down mirror symmetry is broken, and difference in amplitude and phase for upward and downward radiation occurs. The asymmetry parameter Δg determines the broken degree of mirror symmetry and plays a vital role in influencing RA. For directly studying the RA varying as Δg, we look into the RA pseudo-polarization distribution in the synthetic parameter space ($k_x$, Δg) with fixed $k_y$ = 0.

The band structure in the synthetic parameter space is presented in Fig. 2(b) and we focus on the TE$_3$ band. Its corresponding RA pseudo-polarization diagram is plotted in Fig. 2(c). It's noted that a pseudo-polarization vortex exists at ($k_x a/2\pi$, Δg/a)

= (0, 0.113). At the vortex center, the orientation angle becomes ill-defined. However, the orientation angle must be zero for $k_x = 0$, since upward and downward radiation should be identical if they were nonzero due to P symmetry. Consequently, the vortex center can only be a singularity without radiation, which is nothing but a BIC. This can be further substantiated by its field pattern with zero emission (Fig. 2(f)) and diverged quality factor (Supplemental Material [47], Sec. S3). The vortex can be characterized by the topological charge $q = \frac{1}{2\pi} \oint_L d\vec{p} \cdot \nabla_{\vec{p}} \psi(\vec{p})$, where L is a closed loop circling around the vortex center counterclockwise and $\vec{p}$ is a normalized vector in the synthetic parameter space. In this way, the vortex exhibits a topological charge of +1. Moreover, we emphasize that the topological charge is robust against structural parameters such as thickness tg (Supplemental Material [47], Sec. S5).

When we map the loop winding around BIC (pink and light blue line in Fig. 2(c)) onto the Poincaré sphere, as shown in Fig. 2(d), it circles the $S_3$ axis twice, agreeing with a topological charge of +1. In contrast, the loop excluding BIC (yellow line in Fig. 2(c)) doesn't contain the $S_3$ axis, suggesting no any vortex inner the loop. This implies that the BIC vortex enables directionality to be continuously tuned from negative to positive, as the loop containing a vortex must cross the plane at $S_2 = 0$.

Further tuning directionality from -1 to 1 can also be obtained through RA pseudo-polarization diagram. In fact, directionality of -1 and 1, which exactly corresponds to UGR, can be identified as linear pseudo-polarization with orientation angles of 135 ° and 45 °, respectively. Fig. 2(e) plots three characteristic lines determined by RA pseudo-polarization diagram in Fig. 2(c). Lines representing orientation angle of 135° (red) and 45° (blue) originating from the BIC vortex cross with the L line (gray) representing linear pseudo-polarization at ($k_x$ a/2π, Δg/a) = (-0.107, 0.062) and ($k_x$ a/2π, Δg/a) = (0.107, 0.062), thus identifying two UGRs with directionality of -1 and 1. Field patterns in Fig. 2(f) further confirm that distinct unidirectional radiation occurs at these two positions. With the directionality of -1 and 1 identified, then a direct way to adjust directionality from -1 to 1 is to vary parameters along the trajectory linking these two points in the synthetic parameter space, such as changing $k_x$ from -0.107*2π/a to 0.107*2π/a while keeping Δg = 0.062a.

Interestingly, the pseudo-polarization vortex in the synthetic parameter space not only helps to identify the BIC, but also enables us to manipulate the RA from a topological perspective. Owing to conservation law of the topological charge, the BIC vortex can be broken into a pair of C points. Here, we split the BIC vortex by breaking the P symmetry and setting $w_g \neq w_e$, such as shrinking $w_g$ to be 0.674a. As expected, a pair of C points with reversed handedness and identical topological charge of +1/2 are generated in the synthetic parameter space, as shown in Fig. 3(a). The total charge in the synthetic parameter space remains to be +1. Furthermore, accompanied with generated C points, variant RA pseudo-polarization with different orientation and ellipticity angles can be found. Notably, different from real polarization, these diverse pseudo-polarizations actually enable flexible manipulation of RA, including directionality and phase difference, by changing kx and Δg. This can be seen clearly when mapping RA pseudo-polarization onto the Poincaré sphere, as shown in Fig. 3(b), a large area of Poincaré sphere indicating diverse directionality and phase difference is occupied.

Different from P symmetry case, where RA pseudo-polarization is limited to nearly linear, indicating that the phase difference is around 0 or π, as shown in Fig. 2(d), we will show that the broken BIC vortex by breaking P symmetry leads to existing phase difference from π/2 to 3π/2. Since the generated C points locate at the two poles of $S_3$ axis, as shown in Fig. 3(b), it ensures an existing trajectory in the synthetic parameter space, which can be mapped onto the Poincaré sphere as a path linking two poles of $S_3$ axis. As left-handed (upper pole of $S_3$ axis) and right-handed (lower pole of $S_3$ axis) C point represents phase difference of $\delta = \frac{\pi}{2}$ and $\delta = \frac{3\pi}{2}$, respectively, the trajectory linking them means that a continuing modulation of phase difference from π/2 to 3π/2 is guaranteed in the synthetic parameter space. For example, along the trajectory from A to D in Fig. 3(b), the phase difference $\delta$ is varied. As depicted in Fig. 3(c), the point A, B, C, D corresponds to ($k_x$ a/2π, Δg/a) = (3e-4, 0.103), (2e-4, 0.105), (4e-4, 0.109), (3e-4, 0.111) with phase difference of $\delta$ = π/2, 3π/4, 5π/4, 3π/2 respectively. Moreover, their directionality $\eta$ remains to be zero as evidenced by their equal radiation amplitude in upward and downward directions.

In addition, owing to large occupying area of Poincaré sphere, a trajectory indicating equal phase difference $\delta$ but varying directionality $\eta$ can be found as well. For example, along the trajectory from E to H in Fig. 3(b), the phase difference keeps to be $\delta$ = 2π/5. As shown in Fig. 3(d), the point E, F, G, H corresponds to ($k_x$ a/2π, Δg/a) = (-6.2e-3, 0.099), (-2.6e-3, 0.102), (3.7e-3, 0.101), (8.2e-3, 0.098) with directionality of $\eta$ = -0.9, -0.6, 0.6, 0.9, respectively. All of these indicate that by breaking the topological pseudo-polarization vortex in the synthetic parameter space, independent modulation of phase difference and directionality over a large range can be realized, showing a novel way to manipulate the RA of photonic structures.

Finally, we will show that flexible manipulation of RA can actually contribute to design the condition of coherent perfect absorption (CPA) at will by customizing the phase difference of input light. Considering the reciprocal process shown in Fig. 4(a), where light impinges on the bilayer metagratings in opposite directions with phase difference of $\delta_{input}$. For simplicity, we discuss the case of normal incidence. Analysed by temporal coupled mode theory (Supplemental Material [47], Sec. S7. See also references [49-51] therein), it reveals that the coherent absorption satisfies $A = 1 - \frac{1}{2}(1-\eta^2)\left[1-\cos(\delta+\delta_{input})\right]$. This implies that the absorption is determined by two parameters ($\eta, \delta$) of RA. When phase difference of input light $\delta_{input} = -\delta$, CPA happens. In contrast, when $\delta_{input} = -\delta \pm \pi$, $A$ reaches the minimum $A_{min} = \eta^2$. That is to say, by changing $\delta$, we are able to design CPA happening at any $\delta_{input}$.

Induced by C points resulting from the broken vortex, the phase difference $\delta$ can be continuously tuned from π/2 to 3π/2, suggesting that CPA thus can be designed at $\delta_{input}$ from -π/2 to -3π/2. To demonstrate that, we design two types of bilayer metagratings (Supplemental Material [47], Sec. S6) starting from Fig. 3 to match absorption and radiation loss. Here, low intrinsic loss of silicon is considered at 1.078 μm with n = 3.5491+5.55e-5i [48]. As shown in Fig. 4(b), CPA happens at $\delta_{input} = -\frac{\pi}{2}$ when the phase difference of radiation field is $\delta = \frac{\pi}{2}$, while it shifts to $\delta_{input} = -\frac{3\pi}{2}$ for $\delta = \frac{3\pi}{2}$, consistent with the predicted results. When π shift is added, both of cases reaches the minimum absorption. In general, both of absorption shows cosinoidal dependence on $\delta_{input}$, suggesting tunable modulation of absorption by phase of input light. Fig. 4(c) and (d) further demonstrate that all incident light is absorbed due to enhancement of electric field at CPA condition, while the scattering light is nonzero in both side of bilayer metagratings when the condition is violated.

**Conclusions**

In conclusion, we propose a RA pseudo-polarization to investigate arbitrary RA including directionality and phase difference in a unified physical picture. Via the Poincaré sphere representation of RA pseudo-polarization, radiation behaviors of BIC and UGR are clarified as the center and the poles on the sphere, respectively. A robust vortex of RA pseudo-polarization with an integer topological charge is found in the synthetic parameter space of bilayer metagratings. Breaking the vortex into C points with half charges then enables independent modulation of directionality and phase difference over a large range. Furthermore, induced by generated C points, we demonstrate that coherent prefect absorption can be designed at customized phase difference of input light. Our findings provide a new perspective to interpret the topological property of RA for photonic structures, and will further inspire fascinating applications in asymmetric light manipulation, such as one-side detection in optical communication, asymmetric light-light switching and coherent light control for quantum states.

**Acknowledgements**

This work was supported by National Natural Science Foundation of China (Grants No. 62035016, 12274475, 12374364, 12074443), Guangdong Basic and Applied Basic Research Foundation (Grants No. 2023B1515040023, 2023B1515020072), Fundamental Research Funds for the Central Universities, Sun Yat-sen University (23lgbj021, 23ptpy01).

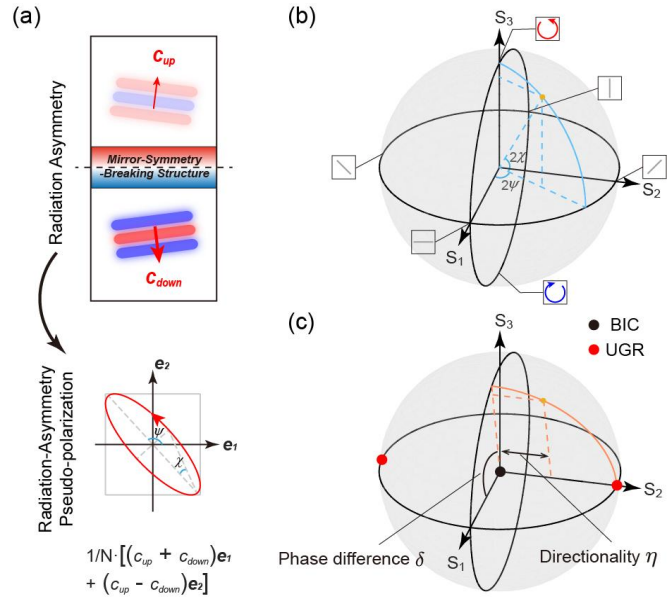

**FIG. 1. Schematic of pseudo-polarization to represent the radiation asymmetry (RA).** (a) Diagram of RA pseudo-polarization $1/N * [(c_{up} + c_{down})e_1 + (c_{up} - c_{down})e_2]$. In general, for mirror-symmetry-breaking structures, it is an ellipse due to amplitude and phase difference of $c_{up}$ and $c_{down}$. (b) RA pseudo-polarization is mapped onto the Poincaré sphere where $\psi$ and $\chi$ are orientation and elliptical angle shown in (a). (c) The behavior of radiation asymmetry is mapped onto the Poincaré sphere, where $\eta$ and $\delta$ are directionality and phase difference between upward and downward radiation field. BIC (black dot) emerges as the center of Poincaré sphere, while UGRs (red dot) locate at the two poles of $S_2$ axis.

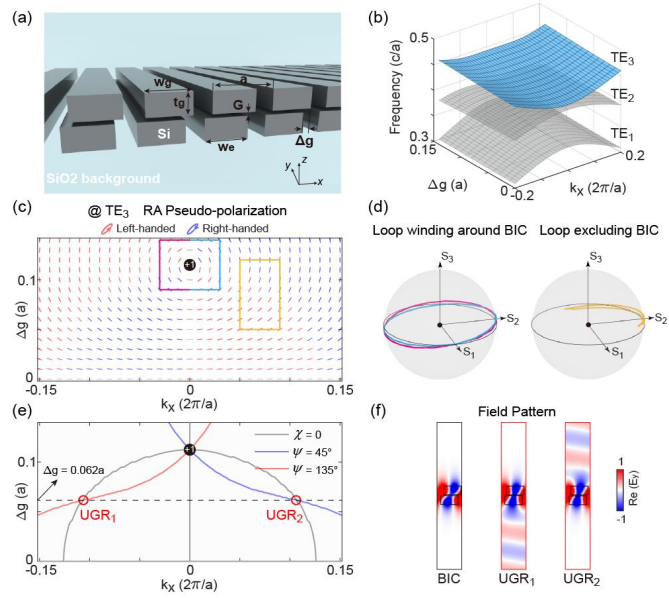

**FIG. 2. RA pseudo-polarization to characterize BIC and UGR in the synthetic parameter space.** (a) Diagram of bilayer metagratings with P-symmetry. (b) Calculated band structure of TE band. (c) RA pseudo-polarization for $TE_3$ band. Red and blue ellipses inside correspond to left-handed and right-handed pseudo-polarizations, respectively. (d) Loop in (c) mapped on the Poincaré sphere, indicating the topological charge of +1 for BIC. (e) Characteristic lines in the synthetic parameter space. Orientation angle of pseudo-polarizations on the red (blue) line is 135° (45°). Gray lines trace out the positions of linear pseudo-polarizations. (f) Field pattern for BIC and UGRs on the $TE_3$ band.

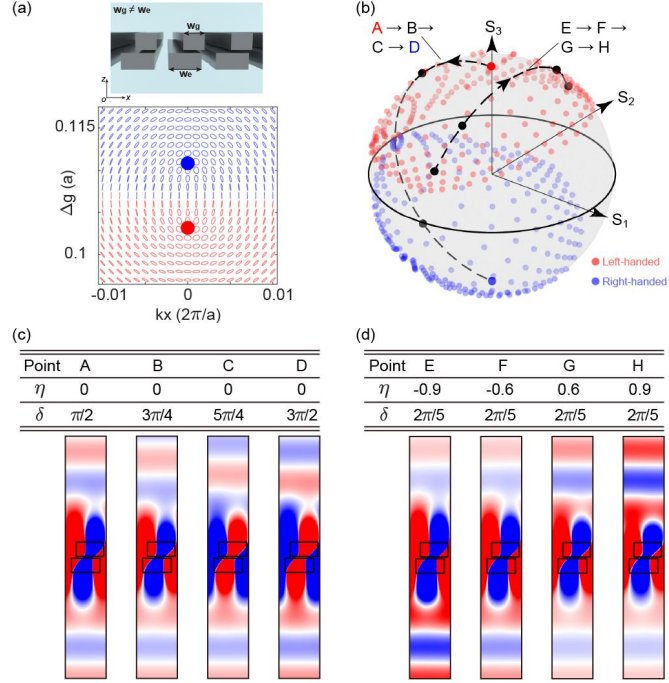

**FIG. 3. Generating diverse directionality and phase difference when breaking the BIC vortex.** (a) (Up) Diagram of bilayer metagratings without P-symmetry ($w_g \neq w_e$). (Down) RA pseudo-polarization for TE$_3$ band suggest left-handed (red) and right-handed (blue) C points with half-integer topological charge are generated after breaking the BIC vortex. (b) RA pseudo-polarizations mapped on the Poincaré sphere. (c) Field patterns for eigen modes along the trajectory in (b). Along the trajectory from A to D, directionality is kept to be 0 while phase difference varies from $\pi/2$ to $3\pi/2$. (d) same as (c) but along the trajectory from E to H, phase difference sustains to be $2\pi/5$, while the directionality varies from -0.9 to 0.9.

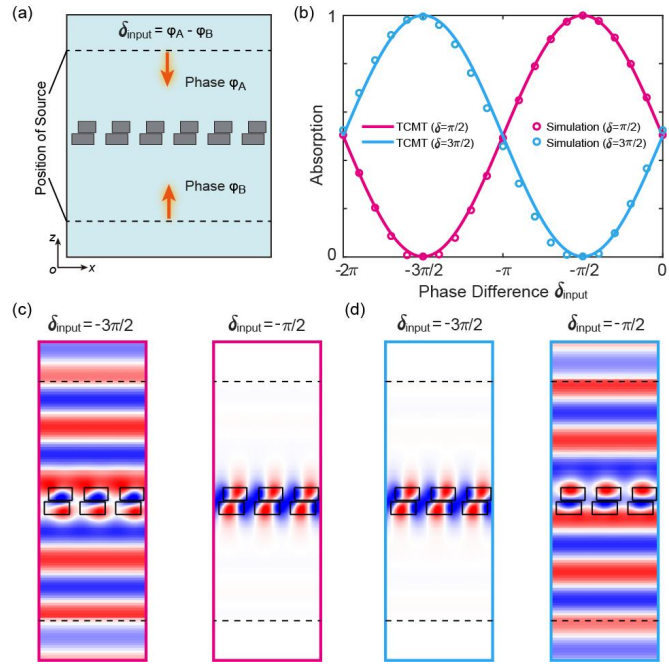

**FIG. 4. Coherent perfect absorption at distinct phase difference.** (a) Schematic of coherent absorption in bilayer metagratings without P symmetry. (b) Absorption varies with the phase difference of input beams $\delta_{input}$. (c, d) Simulated field distribution of bilayer metagratings under two opposite input light sources, whose excited planes are indicated by dashed lines.